\documentstyle[12pt]{article}

\begin{document}

\pagestyle{empty}

\noindent {\small USC-98/HEP-B1\hfill \hfill hep-th/9803188}\newline
{\small \hfill }

{\vskip 0.7cm}

\begin{center}
{\LARGE Gauged Duality, Conformal Symmetry}

{\LARGE and Spacetime with Two Times}$^a${\LARGE \ } \\[0pt]

{\vskip 0.5cm}

{\bf I. Bars}$,${\bf \ C. Deliduman, O. Andreev}{$^b$} {\Large \ \\[0pt]
}

{\vskip 0.4cm}

{\bf Department of Physics and Astronomy}

{\bf University of Southern California}

{\bf Los Angeles, CA 90089-0484}

{\vskip 0.5cm}

{\bf ABSTRACT}
\end{center}

\noindent {\small We construct a duality between several simple physical
systems by showing that they are different aspects of the same quantum
theory. Examples include the free relativistic massless particle and the
hydrogen atom in any number of dimensions. The key is the gauging of the
Sp(2) duality symmetry that treats position and momentum }$(x,p)${\small \
as a doublet in phase space. As a consequence of the gauging, the Minkowski
space-time vectors }$x^\mu ,p^\mu ${\small \ get enlarged by one additional
space-like and one additional time-like dimensions to }$(x^M,p^M).${\small \
A manifest global symmetry SO}$\left( d,2\right) ${\small \ rotates }$%
(x^M,p^M)${\small \ like }$\left( d+2\right) ${\small \ dimensional vectors.
The SO}$\left( d,2\right) ${\small \ symmetry of the parent theory may be
interpreted as the familiar conformal symmetry of quantum field theory in
Minkowski spacetime in one gauge, or as the dynamical symmetry of a totally
different physical system in another gauge. Thanks to the gauge symmetry,
the theory permits various choices of ``time'' which correspond to different
looking Hamiltonians, while avoiding ghosts. Thus we demonstrate that there
is a physical role for a spacetime with two times when taken together with a
gauged duality symmetry that produces appropriate constraints.} \bigskip
\bigskip

\vfill
\hrule width 6.7cm \vskip 2mm

{$^a$ {\small Research partially supported by the U.S. Department of Energy
under grant number DE-FG03-84ER40168, and by the National Research Council
under grant number GAC021197.}}

{$^b$ {\small Permanent address: Landau Institute, Moscow. }}

\vfill\eject

\setcounter{page}1\pagestyle{plain}

\section{Introduction}

The purpose of this paper is to introduce some new points of views on
duality as a gauge symmetry and to connect duality to the concept of a
spacetime with two time-like dimensions. This is an attempt at finding a
physical role for the idea that there may be more than one timelike
dimension to describe our universe at the fundamental level. We will show
that certain familiar physical systems, such as the free massless
relativistic particle, Hydrogen atom, harmonic oscillator, and others, do
fit such a concept, as reported in this paper and in a companion paper \cite
{ibdual}. We will show that these and other apparently different physical
systems correspond to the same quantum Hilbert space characterized by a {\it %
unique} unitary representation of the conformal group SO$\left( d,2\right) $%
. We will argue that the presence of conformal symmetry or dynamical
symmetry in these special cases {\bf is} the evidence for the presence of
two timelike coordinates. The physics looks different because the choice of
``time'' is not unique and hence the Hamiltonians look different, although
they describe the same parent system for which we present an action. These
special physical systems are related to each other by a duality that is a
gauge symmetry. Thanks to the gauge symmetry ghosts are eliminated from the
two time Hilbert space.

Clues for two or more timelike dimensions have been emerging from various
points of view, including the brane-scan \cite{duff}, the structure of
extended supersymmetry of p-branes \cite{ibtokyo}, extensions of M-theory 
\cite{mtheory} to F-theory \cite{ftheory} and S-theory \cite{stheory}-\cite
{14d}, (1,2) strings \cite{martinec}, 12D super Yang-Mills \& supergravity
theories in backgrounds of constant lightlike vectors\cite{sezgin}, and
finally the discovery of models of multi-superparticles that are fully
covariant in (10,2) and (11,3) dimensions \cite{ibkounnas}-\cite{sezrudy}.

Two or more timelike dimensions are possible only with appropriate gauge
symmetry and constraints that reduce the theory to an effective theory with
a single timelike dimension and no ghosts. The gauged Sp$\left( 2\right) $
duality symmetry suggested here is an evolution of the local bosonic
symmetry introduced in \cite{ibkounnas}-\cite{sparticles} for the same
purpose. The difference is that we apply the concept to the phase space
doublet $(X^M,P^M)$ for a single particle rather than to a multiplet of the
positions of several particles $(X_1^M,X_2^M,\cdots )$. We suggest an action
principle in phase space, including invariant interactions with background
fields, with and without supersymmetry.

We have suggestively named our local symplectic symmetry Sp$\left( 2\right) $
``duality'' because we see signs that our duality is related to the
generalized concept of electric-magnetic duality in super Yang-Mills
theories and M-theory. However, this connection remains to be established by
further detailed study.

\section{Gauging duality}

The quantization rules of quantum mechanics are symmetric under the
interchange of coordinates and momenta. This is known as the symplectic
symmetry Sp$(2)$ that transforms $(x,p)$ as a doublet. Maxwell's equations
for electricity and magnetism are symmetric under the interchange of
electricity and magnetism in the absence of sources. The electric and
magnetic fields are generalized coordinates and momenta. In the presence of
particles with quantized electric and magnetic charges the symmetry is a
discrete version of Sp$\left( 2\right) $. This symmetry, known as
``electric-magnetic duality'', is apparently broken in our part of the
universe by the absence of magnetic monopoles and dyons. The idea of
electric-magnetic duality symmetry has been generalized in recent
non-perturbative studies of supersymmetric field theory \cite{seibwitt} and
string theory \cite{hulltown}, which are now believed to be only some aspect
of a larger, duality invariant, mysterious theory (M-theory, F-theory,
S-theory, U-theory, $\cdots $). In the context of the mysterious theory,
``duality'', which is a much larger symmetry than Sp(2), but containing it,
is believed to be a gauge symmetry.

In this paper we study an elementary system with local continuous Sp$\left(
2\right) $ duality symmetry. We start by reformulating the worldline
description of the standard free massless relativistic point particle by
gauging the Sp(2) duality symmetry. What we find in doing so is a more
general theory capable of describing not only the free particle but other
physical systems dual to it, such as the hydrogen atom, harmonic oscillator,
and others.

To remove the distinction between $x$ and $p$ we will rename them $%
X_1^M\equiv X^M$ and $X_2^M\equiv P^M$ and define the doublet $X_i^M=\left(
X_1^M,X_2^M\right) .$ The local Sp$\left( 2\right) $ acts as follows 
\begin{equation}
\delta _\omega X_i^M\left( \tau \right) =\varepsilon _{ik}\omega ^{kl}\left(
\tau \right) X_l^M\left( \tau \right) .  \label{doublet}
\end{equation}
Here $\omega ^{ij}\left( \tau \right) =\omega ^{ji}\left( \tau \right) $ is
a symmetric matrix containing three local parameters, and $\varepsilon _{ij}$
is the Levi-Civita symbol that is invariant under Sp$\left( 2,R\right) $ and
serves to raise or lower indices. We also introduce an Sp$\left( 2,R\right) $
gauge field $A^{ij}\left( \tau \right) $ which is symmetric in $(ij)$ which
transforms in the standard way 
\begin{equation}
\delta _\omega A^{ij}=\partial _\tau \omega ^{ij}+\omega ^{ik}\varepsilon
_{kl}A^{lj}+\omega ^{jk}\varepsilon _{kl}A^{il}.
\end{equation}
The covariant derivative is 
\begin{equation}
D_\tau X_i^M=\partial _\tau X_i^M-\varepsilon _{ik}A^{kl}X_l^M.
\label{covder}
\end{equation}
An action that is invariant under this gauge symmetry is 
\begin{eqnarray}
S_0 &=&\frac 12\int_0^Td\tau \left( D_\tau X_i^M\right) \varepsilon
^{ij}X_j^N\eta _{MN}  \label{action} \\
&=&\int_0^Td\tau \left( \partial _\tau X_1^MX_2^N-\frac
12A^{ij}X_i^MX_j^N\right) \eta _{MN},  \nonumber
\end{eqnarray}
Here $\eta _{MN}$ is a flat metric in $d+2$ dimensions and a total
derivative has been dropped in rewriting the first term. The signature of
the metric $\eta _{MN}$ is not specified at this stage, but we will see that
it will be {\it imposed} on us that it must have signature for two timelike
dimensions. From the second form of the action one may identify the
canonical conjugates as $X_1^M=X^M$ and $\partial S/\partial \dot{X}%
_1^M=X_2^M=P^M$, so that the action is consistent with the idea that $%
(X_1^M,X_2^M)$ is the doublet $\left( X^M,P^M\right) $ rather than
describing two particles.

If instead of the full Sp$\left( 2\right) $ group we had gauged a triangular
abelian subgroup containing only $\omega ^{22}\left( \tau \right) ,$ and
kept only the gauge potential $A^{22}\left( \tau \right) $, then the
resulting action would have been the free massless particle action in the
first order formalism, with $\eta _{\mu \nu }$ the standard Minkowski
metric. Thus $\omega ^{22}$ is closely related to $\tau $ reparametrization
invariance, but $\omega ^{12},\omega ^{11}$ are new local symmetry
parameters that permit the removal of redundant gauge degrees of freedom. In
the presence of the gauge degrees of freedom we are able to see the
structure of duality and the role it plays in exhibiting higher symmetries
in higher dimensions.

In addition to the local Sp$\left( 2,R\right) $ symmetry there is a manifest
global symmetry SO$\left( d,2\right) $ (assuming signature $\left(
d,2\right) $) acting on the space-time $X_{i}^{M}$ with $d$-spacelike and $2$%
-timelike dimensions labelled by the index $M$. This symmetry contains the $%
d $-dimensional Poincar\'{e} symmetry ISO$\left( d-1,1\right) $ as a
subgroup, but there is no translation symmetry in $d+2$ dimensions. Using
No\"{e}ther's theorem one finds the generators of the symmetry SO$\left(
d,2\right) $ 
\begin{equation}
L^{MN}=\varepsilon ^{ij}X_{i}^{M}X_{j}^{N}=X^{M}P^{N}-X^{N}P^{M}.
\label{generators}
\end{equation}
They are manifestly {\it gauge invariant} under the local Sp$\left(
2,R\right) $ transformations.

Supersymmetry on the worldline is introduced using the Neveu-Schwarz
approach but only for zero modes. To do so, phase space is enlarged by the
addition of fermionic degrees of freedom $\psi ^M\left( \tau \right) $ which
are their own canonical conjugates (i.e. they form a Clifford algebra when
quantized). The Sp$\left( 2\right) $ doublet is enlarged to an OSp$\left(
1/2\right) $ triplet $(\psi ^M,X_1^M,X_2^M)$ and the supergroup OSp$\left(
1/2\right) $ is gauged by adding two fermionic gauge potentials $F_i$ in
addition to the three bosonic gauge potentials $A^{ij}$. The action is the
direct generalization of (\ref{action}) to a gauge theory based on OSp$%
\left( 1/2\right) $. In a particular gauge, the degrees of freedom reduce
correctly to the free Dirac particle in Minkowski space. This scheme can be
enlarged to $N$ supersymmetries by gauging OSp$\left( N/2\right) $. Like the
bosonic case, the supersymmetric case also has multiple physical sectors as
seen from the point of view of various gauge choices for ``time''. The
supersymmetric case will be discussed in more detail in another paper \cite
{osp12}.

Interactions with gravitational fields $G_{MN}\left( X_1,X_2\right) $ and
gauge fields $A_j^N\left( X_1,X_2\right) $ in a way that respects the Sp$%
\left( 2\right) $ duality symmetry are possible (of course, also in the
supersymmetric case) 
\begin{equation}
S_{G,A}=\frac 12\int_0^Td\tau \left[ 
\begin{array}{c}
\left( D_\tau X_i^M\right) \varepsilon ^{ij}X_j^N\,\,G_{MN}\left(
X_1,X_2\right) \\ 
+\left( D_\tau X_i^M\right) \varepsilon ^{ij}A_{jN}\left( X_1,X_2\right)
\,.\,
\end{array}
\right]  \label{sga}
\end{equation}
$G_{MN}$ is a scalar under Sp$\left( 2\right) $ and a symmetric traceless
tensor in $d+2$ dimensions. Similarly $A_j^M$ is a doublet under Sp$\left(
2\right) $ and a vector in $d+2$ dimensions. It is tempting to suggest that
the Sp$\left( 2\right) $ doublet of electromagnetic fields $A_j^M$ are
related to the electric-magnetic dual potentials of Maxwell's theory and its
Yang-Mills generalizations. For the local invariance to hold, there must be
restrictions on the functional forms of both $G_{MN}\left( X_1,X_2\right) $
and $A_j^N\left( X_1,X_2\right) $ since the arguments $\left( X_1,X_2\right) 
$ also transform under Sp$\left( 2\right) $. These amount to a set of
differential equations that restrict the functional forms of $G_{MN}\left(
X_1,X_2\right) $ and $A_j^N\left( X_1,X_2\right) $. One automatic solution
is to take any functions $G_{MN}\left( L\right) ,$ $A_j^N\left( L\right) $
where $L^{MN}$ is the gauge invariant combination of $\left( X_1,X_2\right) $
given in (\ref{generators})$.$ In the presence of the background fields the
global symmetry SO$\left( d,2\right) $ is replaced by the Killing symmetries
of the background fields. We see that, for consistency with the local
symmetry, gravity and gauge interactions are more conveniently expressed in
terms of bi-local fields $G_{MN}\left( X_1,X_2\right) $ and $A_j^N\left(
X_1,X_2\right) $ in $d+2$ dimensions. Bi-local fields were advocated in \cite
{stheory} as a means of extending supergravity and super Yang-Mills theory
to (10,2) dimensions based on clues from the BPS solutions of extended
supersymmetry.

We refer to the forms of the actions $S_0,S_{G,A}$ above as the first order
formalism. Although not necessary, a second order formalism is obtained if $%
X_2^M$ is integrated out in the path integral (or eliminated
semi-classically through one of the equations of motion). Eliminating $X_2^M$
is not easy for the interacting case, but for the free action $S_0$ the
result is 
\begin{equation}
S_0=\int d\tau \,\left[ \frac 1{2A^{22}}\left( \partial _\tau
X^M-A^{12}X^M\right) ^2-\frac{A^{11}}2\,X\cdot X\right] ,  \label{confaction}
\end{equation}
This form of the action may be thought of as ``conformal gravity'' on the
worldline, with the conformal group SO$\left( 1,2\right) =$Sp$\left(
2\right) $.

In this paper we will mainly analyze the simplest case $S_0$. The
configuration space version of $S_0$ (\ref{confaction}) was previously
obtained with different reasoning and motivation \cite{marnelius}\footnote{%
We thank K. Pilch for discovering this reference at the time of publication.}%
, and without the concept of duality. Our solutions to both the classical
and quantum problems go well beyond previous discussion of this system \cite
{siegel}-\cite{followup}. More importantly, our interpretation of the system
and its scope as a theory for duality and two times, and the applications to
physical situations are new.

\section{Classical solutions and dual sectors}

The equations of motion for $(X_1,X_2)$ in the case of $S_0$ are 
\begin{equation}
\left( 
\begin{array}{c}
\partial _\tau X^M \\ 
\partial _\tau P^M
\end{array}
\right) =\left( 
\begin{array}{cc}
A_{12} & A_{22} \\ 
-A_{11} & -A_{12}
\end{array}
\right) \left( 
\begin{array}{c}
X^M \\ 
P^M
\end{array}
\right) .  \label{motion}
\end{equation}
In addition, the equations of motion for the $A_{ij}$ produces the
constraints 
\begin{equation}
X\cdot X=0,\quad X\cdot P=0,\quad P\cdot P=0.  \label{constraints}
\end{equation}
At least two timelike dimensions are required to obtain non-trivial
solutions to the constraints \cite{ibkounnas}, and our gauge symmetry does
not allow more than two timelike dimensions without running into problems
with ghosts. Thus our system exists physically only with the signature $%
\left( d,2\right) $.

To show that the massless Minkowski particle is one of the classical
solutions of our system, we may choose the gauge $A^{12}=A^{11}=0$ and $%
A^{22}=1$, solve the equations $X^M=Q^M+P^M\tau $, and obtain the
constraints $Q^2=P^2=Q\cdot P=0.$ There is a remaining gauge symmetry 
\begin{eqnarray}
\omega ^{11}\left( \tau \right) &=&\omega _0^{11},\quad \omega ^{12}\left(
\tau \right) =-\omega _0^{11}\tau +\omega _0^{12}, \\
\omega ^{22}\left( \tau \right) &=&\omega _0^{11}\tau ^2-2\omega _0^{12}\tau
+\omega _0^{22},  \nonumber
\end{eqnarray}
where $\omega _0^{ij}$ are $\tau $-independent constants. Next define the
basis $Q^M=(Q^{+^{\prime }},Q^{-^{\prime }},q^\mu ),$ $P^M=(P^{+^{\prime
}},P^{-^{\prime }},p^\mu )$, where $\pm ^{\prime }$ indicate a lightcone
type basis for the extra (1,1) dimensions with metric $\eta ^{+^{\prime
}-^{\prime }}=-1$. Using two parameters of the remaining gauge freedom
choose $Q^{+^{\prime }}=1$, $P^{+^{\prime }}=0$, and solve the two
constraints $Q^2=Q\cdot P=0$, so that the solution takes the form 
\begin{eqnarray}
X^{+^{\prime }}\left( \tau \right) &=&1,\quad X^{-^{\prime }}\left( \tau
\right) =\frac{q^2}2+q\cdot p\,\tau \,,  \nonumber \\
X^\mu \left( \tau \right) &=&q^\mu +p^\mu \tau ,\quad p^2=0\quad massless.
\label{freemassless}
\end{eqnarray}
There remains one free gauge parameter $\omega _0^{22}$ and one constraint $%
P^2=p^2=0$, which is also what follows from $\tau $ reparametrizations on
the worldline. The motion in $d$-dimensional Minkowski subspace $x^\mu
\left( \tau \right) $ is the same as the standard massless particle.
Furthermore, the motion in the remaining two coordinates $X^{+^{\prime
}},X^{-^{\prime }}$ is fully determined by the position and momentum $(q^\mu
,p^\mu )$ in Minkowski space.

The free massless particle is not the only classical solution. For example,
in the gauge $A^{12}=0$, $A^{11}=A^{22}=\omega $ the solution is 
\begin{eqnarray}
X_M &=&a_M\,e^{i\omega \tau }+a_M^{\dagger }\,e^{-i\omega \tau }, \\
a\cdot a &=&a^{\dagger }\cdot a=a^{\dagger }\cdot a^{\dagger }=0.  \nonumber
\end{eqnarray}
This is an oscillatory motion with a different physical interpretation than
the free relativistic particle. As we will see, our system has dual sectors
that include the H-atom and harmonic oscillator, which evidently are
periodic systems. Some previously known solutions include a massive particle
in Minkowski space \cite{marnelius}, a massless particle in deSitter space 
\cite{marnelius}, etc.. Thus, there are classical solutions of the same
system with various physical meanings.

What is going on is that choosing ``time'' is tricky in our system since
there is more than one timelike dimension. The dynamics of the system is
arranged to evolve according to some gauge choice of ``time'' which is not
unique in the system. For each such choice there is a corresponding
canonical conjugate Hamiltonian which looks like different physics. However,
there really is one single overall theory that follows from our action. It
has various physical interpretations that are dual to each other, where
duality is the Sp$\left( 2\right) $ gauge symmetry that we have introduced.
Under Sp$\left( 2\right) $ transformations every classical solution which
has a different physical interpretation in some gauge can be mapped to the
free massless particle by a gauge transformation and a different choice of
``time''.

There is a gauge invariant way to characterize the overall system at the
classical as well as quantum levels. The SO$\left( d,2\right) $ global
symmetry generators $L^{MN}$ are gauge invariant, as well as constants of
motion with respect to the ``time'' $\tau $. Using the constraints, it is
straightforward to compute that all the Casimirs of SO$\left( d,2\right) $
vanish at the classical level 
\begin{equation}
C_n\left( SO\left( d,2\right) \right) =\frac{-1}{n!}Tr\left( L\right)
^n=0,\quad classical.
\end{equation}
For a non-compact group such a representation is non-trivial. For example
the free particle is such a representation. This can be verified by
inserting the free particle gauge of eq.(\ref{freemassless}) into (\ref
{generators}). As we will see, the Casimirs $C_n$ will not all be zero at
the quantum level, when ordering of operators are taken into account. We
will find very specific values in the quantum gauge invariant sector, in
particular $C_2\left( SO\left( d,2\right) \right) =1-d^2/4$. Both at the
classical and quantum levels, the Casimir invariants specify a {\it unique}
unitary representation of SO$\left( d,2\right) $ which fully characterizes
the gauge invariant physical space of the system. This approach does not
involve a choice of ``time'' or Hamiltonian or effective Lagrangian in a
fixed gauge.

Having realized this important observation one may now understand more
generally that in a special gauge we find a rather non-trivial classical and
quantum solution of our system, namely the Hydrogen atom in any dimension
(the non-relativistic central force problem with the $1/r$ potential). The
essential reason for its existence is that all the levels of the H-atom
taken together form a single irreducible representation of the conformal
group SO$\left( d,2\right) ,$ in accord with the observation above. In fact,
the representation is precisely the unique one that emerges from quantum
ordering (next section), with specific values of the Casimirs. It was known
that the H-atom in three dimensions ($d-1=3)$ forms a single irreducible
representation of SO(4,2) \cite{barut}. The well known SO$\left( 4\right) $
symmetry is the subgroup of SO$\left( 4,2\right) $. This solution will be
fully explained and generalized to any dimension at the classical and
quantum levels in a separate paper \cite{ibdual} (with quantum ordering and
other technical aspects that differ from the old literature \cite{barut}).
It will also be shown there that the harmonic oscillator in $\left(
d-2\right) $ dimensions, with its mass equal to a lightcone momentum in an
additional dimension, is also a solution of the system. As for all
solutions, the H-atom or harmonic oscillator are Sp$\left( 2\right) $ dual
to the free massless relativistic particle!

To close this section we provide a general parametrization of classical
solutions in any gauge. We take advantage of the fact that the SO$\left(
d,2\right) $ generators are constants of motion $\partial _\tau L^{MN}=0$
with respect to the ``time'' $\tau $. A general classical solution in any
gauge may be given in various bases $M=(+^{\prime },-^{\prime },\mu )$, $%
\,M=(0^{\prime },1^{\prime },\mu )$, $\,M=\left( 0^{\prime },0,I\right) $.
The first is a lightcone type basis in the extra dimensions $X^{+^{\prime
}}=\left( X^{0^{\prime }}+X^{1^{\prime }}\right) ,\,\,X^{-^{\prime }}=\frac
12\left( X^{0^{\prime }}-X^{1^{\prime }}\right) $, and the last
distinguishes the two timelike coordinates from the spacelike ones. The
first two are covariant under SO$\left( 1,1\right) \otimes S\left(
d-1,1\right) $ and the last is covariant under SO$\left( 2\right) \otimes $SO%
$\left( d\right) .$ The general solution is 
\begin{equation}
\begin{array}{cccc}
M\,\,\,\,= & [\,+^{\prime }, & -^{\prime }, & \,\,\,\,\,\,\,\,\,\,\,\mu
\,\,\,\,\,\,\,\,\,\,\,\,\,\,\,\,\,\,\,\,\,\,] \\ 
X^M= & [\,\,a\,\,, & b{\bf ,} & \frac{-aL^{-^{\prime }\mu }+bL^{+^{\prime
}\mu }}{ad-bc}\,\,] \\ 
P^M= & [\,\,\,c, & d, & \,\frac{-cL^{-^{\prime }\mu }+dL^{+^{\prime }\mu }}{%
ad-bc}\,\,\,{\bf ]}
\end{array}
\label{lcone}
\end{equation}
with 
\begin{equation}
\left( 
\begin{array}{cc}
A_{12} & A_{22} \\ 
-A_{11} & -A_{12}
\end{array}
\right) =\left( 
\begin{array}{cc}
\partial _\tau a & \partial _\tau b \\ 
\partial _\tau c & \partial _\tau d
\end{array}
\right) \left( 
\begin{array}{cc}
a & b \\ 
c & d
\end{array}
\right) ^{-1},
\end{equation}
where the matrix $\left( a\left( \tau \right) ,b\left( \tau \right) ,c\left(
\tau \right) ,d\left( \tau \right) \right) $ is a group element of GL$\left(
2,R\right) $. It can be checked that by inserting this form into (\ref
{generators}) that the constants $L^{\pm ^{\prime }\mu }$ that appear in (%
\ref{lcone}) are consistent with their definitions. Another constant of
motion is the determinant of the matrix 
\begin{equation}
L^{+^{\prime }-^{\prime }}=ad-bc.
\end{equation}
So, effectively the local gauge group is Sp$\left( 2\right) $ as
parametrized by $\left( a,b,c,d\right) $. The remaining generators $L^{\mu
\nu }$ which are also constants of motion, are now written in terms of the
constants $L^{+^{\prime }-^{\prime }},$ $L^{\pm ^{\prime }\mu }$%
\begin{equation}
L^{\mu \nu }=X^\mu P^\nu -X^\nu P^\mu =\frac 1{L^{+^{\prime }-^{\prime
}}}\left( L^{+^{\prime }\mu }L^{-^{\prime }\nu }-L^{-^{\prime }\mu
}L^{+^{\prime }\nu }\right) .
\end{equation}
We may forget completely about the gauge potentials $A^{ij}$ and concentrate
instead on the local group element $\left( a,b,c,d\right) $ and the global
symmetry generators $L^{MN}$. The constraints (\ref{constraints}) become
conditions to be satisfied by the $L^{\pm ^{\prime }\mu },L^{+^{\prime
}-^{\prime }},L^{\mu \nu }$ without any condition on the group element 
\begin{eqnarray}
L^{+^{\prime }\mu }L^{+^{\prime }\nu }\eta _{\mu \nu } &=&L^{-^{\prime }\mu
}L^{-^{\prime }\nu }\eta _{\mu \nu }=0,  \nonumber \\
\frac 12\left( L^{\mu \nu }\right) ^2 &=&L^{+^{\prime }\mu }L^{-^{\prime
}\nu }\eta _{\mu \nu }=-\left( L^{+^{\prime }-^{\prime }}\right) ^2.
\label{lcon}
\end{eqnarray}
With these conditions the Casimir for SO$\left( d,2\right) $ becomes 
\begin{equation}
C_2=\frac 12\left( L^{MN}\right) ^2=0,  \label{L2class}
\end{equation}
at the classical level, and similarly for all higher Casimirs. But we will
see below that in the quantum theory, when we watch the orders of the
operators, the quadratic Casimir will be $C_2=1-\frac{d^2}4.$ Similarly all
higher Casimirs of SO$\left( d,2\right) $ vanish at the classical level, but
not at the quantum level.

The same arguments may be repeated in the other bases. For example, in the
basis $M=\left( 0^{\prime },0,I\right) $ we have 
\begin{equation}
\begin{array}{cccc}
M\,\,\,\,= & [\,0^{\prime }, & 0, & \,\,\,\,\,\,\,\,\,\,\,I\,\,\,\,\,\,\,\,%
\,\,\,\,\,\,\,\,\,\,\,\,\,\,] \\ 
X^M= & [\,\,a\,\,, & b{\bf ,} & \frac{-aL^{0I}+bL^{0^{\prime }I}}{ad-bc}\,\,]
\\ 
P^M= & [\,\,\,c, & d, & \,\frac{-cL^{0I}+dL^{0^{\prime }I}}{ad-bc}\,\,\,{\bf %
]}
\end{array}
\label{timeli}
\end{equation}
with 
\begin{eqnarray}
L^{0^{\prime }0} &=&ad-bc\,\,,  \nonumber \\
L^{IJ} &=&\frac 1{L^{0^{\prime }0}}\left( L^{0^{\prime
}I}L^{0J}-L^{0^{\prime }J}L^{0I}\right) \,\,,
\end{eqnarray}
and 
\begin{equation}
\left( L^{0I}\right) ^2=\left( L^{0^{\prime }I}\right) ^2=-\frac 12\left(
L^{IJ}\right) ^2=-\left( L^{0^{\prime }0}\right) ^2,  \label{timel}
\end{equation}
so that again (\ref{L2class}), and the same conditions on the higher order
Casimirs hold.

\section{Quantum theory}

In any gauge the naive quantum rules that follows from the action $S_0$ are $%
\left[ X_i^M,X_j^N\right] =i\varepsilon _{ij}\eta ^{MN}.$ These are subject
to the constraints $X_i\cdot X_j=0.$ We will rewrite these in any gauge as 
\begin{equation}
\left[ X^M,P^N\right] =i\eta ^{MN},\quad X^2=P^2=X\cdot P=0.
\end{equation}
As usual one may approach the problem of quantization in a covariant
formalism or in a non-covariant formalism.

In a covariant formalism one may apply the constraints on states constructed
in a Hilbert space that obeys the naive quantization rules above. This
approach would be manifestly covariant under both the duality symmetry Sp$%
\left( 2,R\right) $ as well as the SO$\left( d,2\right) $ symmetry. But is
does not seem to give direct insight into the physical content of the theory
since ``time'' or ``Hamiltonian'' is not specified. In this paper we will
obtain one crucial result on the values of the Casimirs $C_n\left( SO\left(
d,2\right) \right) $ that follow from covariant quantization.

In a non-covariant formalism both the duality symmetry as well as the
manifest SO$\left( d,2\right) $ symmetry is broken by the choice of gauges
and solution of the constraints. One must then verify that the quantization
procedure respects the gauge invariant algebra of the global SO$\left(
d,2\right) $ generators $L^{MN}$ in eq.(\ref{generators}). In the gauge
fixed formalism these generators incorporate the naive global transformation
on the $d+2$ space-time coordinates as well as the duality transformations Sp%
$\left( 2,R\right) .$ This is because after an SO$\left( d,2\right) $
transformation one goes out of the gauge slice, and a gauge transformation
must be applied to go back to the gauge slice. Thus, the details of the SO$%
\left( d,2\right) $ conformal generators in the fixed gauge provide the
information on the duality transformations. In a fixed gauge some of the $%
L^{MN}$ require normal ordering of the canonical degrees of freedom and
therefore there are anomaly coefficients. The closure of the algebra can fix
some of these coefficients, but it turns out this is not so in every gauge.
It turns out that imposing the eigenvalues of the Casimirs $C_n$ obtained in
the covariant quantization must be used to fully determine the anomaly
coefficients. In particular for the Hydrogen atom this additional constraint
is needed. In this paper we will treat only the free particle in two
different gauges and verify that we have the correct representation.

\subsection{SO$\left( d,2\right) $ and Sp$\left( 2\right) $ covariant
quantization}

The hermitian quantum generators of Sp$\left( 2,R\right) $ are 
\begin{eqnarray}
J_0 &=&\frac 14\left( P^2+X^2\right) ,\quad J_1=\frac 14\left(
P^2-X^2\right) ,  \label{sp2} \\
J_2 &=&\frac 14\left( X\cdot P+P\cdot X\right) .
\end{eqnarray}
The Lie algebra that follows from the quantum rules is 
\begin{equation}
\left[ J_0,J_1\right] =iJ_2,\quad \left[ J_0,J_2\right] =-iJ_1,\quad \left[
J_1,J_2\right] =-iJ_0.
\end{equation}
The quadratic Casimir operator $C_2(Sp\left( 2\right) )=J_0^2-J_1^2-J_2^2$
takes the hermitian form (watching the orders of operators) 
\begin{equation}
C_2(Sp\left( 2\right) )=\frac 14\left[ X^MP^2X_M-\left( X\cdot P\right)
\left( P\cdot X\right) +\frac{d^2}4-1\right]  \label{sl2casimir}
\end{equation}
where the constant term arises from re-ordering the operators $%
(d+2)^2-4\left( d+2\right) =d^2-4.$ The gauge invariant SO$\left( d,2\right) 
$ Lorentz generators given in eq.(\ref{generators}) are used to compute the
quadratic Casimir operator for SO$\left( d,2\right) $. One finds that the
quadratic Casimirs of the two groups are related 
\begin{equation}
C_2\left( SO\left( d,2\right) \right) =\frac 12L_{MN}L^{MN}=\left[
C_2(Sp\left( 2\right) )+1-\frac{d^2}4\right] ,  \label{c2}
\end{equation}
where $C_2(Sp\left( 2\right) )$ is given by eq.(\ref{sl2casimir}). Since $%
L_{MN}$ is gauge invariant, both $C_2\left( SO\left( d,2\right) \right) $
and $C_2(Sp\left( 2\right) )$ must have the same spectrum in any
quantization scheme in any gauge.

We will describe the general properties of the covariant Hilbert space we
should find. The ``physical'' states form a subset of the Hilbert space for
which the matrix elements of Sp$\left( 2,R\right) $ generators vanish weakly 
\begin{equation}
<phys|J_{0,1,2}|phys^{\prime }>\sim 0.  \label{physcond}
\end{equation}
For SL$\left( 2,R\right) =$Sp$\left( 2,R\right) $ all the unitary
representations are labelled by $|jm>$. Within this space the singlet state $%
C_2(Sp\left( 2\right) )=j(j+1)=0$ and $m=0$, satisfy the physical
requirements. This is the module with only one state from the point of view
of Sp$\left( 2,R\right) $. However, there can be an infinite number of such
gauge invariant states which are classified by the global symmetry SO$\left(
d,2\right) $. This must be the case since we already know that there is a
non-trivial solution of the constraints in the classical limit when the
signature of $\eta ^{MN}$ is $\left( d,2\right) $\footnote{%
There is another trivial state that satisfies the physical state conditions
with some modification in the weak condition. This is the Fock vacuum if one
uses a harmonic oscillator representation with $X_M=(a_M+a_M^{\dagger })/%
\sqrt{2}$ and $P_M=(a_M-a_M^{\dagger })/\sqrt{2}i$. Taking into account
operator ordering, then one finds $J_0=\frac 12a^{\dagger }\cdot a+\frac
14(d+2)$ and compute that the Fock vacuum has $j(j+1)=-1+d^2/4$ and $%
m_0=\frac 14(d+2)$. The physical state condition gets modified to $J_0=\frac
14(d+2)$ instead of zero. This state is the lowest state of the non-trivial
discrete series representation of Sp$\left( 2\right) $. However, it is the
trivial singlet state from the point of view of SO$\left( d,2\right) $ since 
$L_{MN}=a_M^{\dagger }a_N-a_N^{\dagger }a_M$ annihilates it. This is the
only state that would exist in the theory if the signature were $\left(
d+2,0\right) $ or $\left( d+1,1\right) $.}. Thus, we have argued that for
non-trivial states we must have 
\begin{equation}
C_2\left( SO\left( d,2\right) \right) =1-\frac{d^2}4,\quad C_2(Sp\left(
2\right) )=0.\,
\end{equation}
This will be confirmed by the non-covariant quantization below. To compute
the eigenvalues of all the Casimir operators $C_n$ we use the same approach.
We find that all $C_n$ at the quantum level can first be written in terms of 
$C_2(Sp\left( 2\right) )$ plus normal ordering constants that depend on $d$.
Once the general expression is obtained we set $C_2(Sp\left( 2\right) )=0$
and obtain the eigenvalues of $C_n(SO(d,2))$ for the gauge invariant states.
This procedure uniquely determines the physical space content of our theory
as a {\it unique unitary representation} of the conformal group SO$\left(
d,2\right) $. We only need the quadratic Casimir in the present paper.

Although we have identified the physical representation of Sp$\left(
2,R\right) $ and SO$\left( d,2\right) $, building explicitly the Sp$\left(
2,R\right) $ and SO$\left( d,2\right) $ fully covariant Hilbert space in
terms of the covariant canonical variables $X^M,P^M$ remains as an open
problem. For a physical interpretation this is desirable. A natural approach
to study the general problem covariantly is in terms of bi-local fields $%
\phi (X_1^M,X_2^M)$. Recall that bi-local fields are also relevant as
background fields in the general action $S_{G,A}$.

\subsection{Fully gauge fixed quantization}

In the non-covariant approach we choose a gauge and solve all the
constraints at the classical level, and then quantize the remaining degrees
of freedom. The advantage of this approach is that unitarity is manifest and
we work directly with the physical states. The disadvantage is that by
choosing a gauge we hide the duality properties. We will discuss here only
the free massless particle interpretation of the Hilbert space. In another
paper we will show that the same Hilbert space is dual the H-atom and
harmonic oscillator. We fix three gauges that make evident the free particle
interpretation as in the classical solution (\ref{freemassless}) $%
X^{+^{\prime }}=1,P^{+^{\prime }}=0,X^{+}=p^{+}\tau $. Since we will express
the commutation rules at $\tau =0$, we have, in a lightcone basis $M=\left(
+^{\prime },-^{\prime },+,-,i\right) $ $\,$ 
\begin{equation}
X^M=\left( 1,x^{-^{\prime }};0,x^{-};{\bf x}^i\right) ,\quad P^M=\left(
0,p^{-^{\prime }};p^{+},p^{-},{\bf p}^i\right)  \label{gauge1}
\end{equation}
where the transverse vectors ${\bf x}^i,{\bf p}^i$ are in $\left( d-2\right) 
$ dimensions. Inserting this form in the constraints gives 
\begin{equation}
x^{-^{\prime }}=\frac{{\bf x}^2}2,\quad p^{-^{\prime }}=\left( {\bf x}\cdot 
{\bf p}-x^{-}p^{+}\right) ,\quad p^{-}=\frac{{\bf p}^2}{2p^{+}}.
\label{gauge11}
\end{equation}
where we have used $\eta ^{+^{\prime }-^{\prime }}=\eta ^{+-}=-1$. The
canonical pairs are 
\begin{eqnarray}
&&\left[ {\bf x,p}\right] ,\quad \left[ x^{-},p^{+}\right] ,\quad
[x^{+}=0,\,\,p^{-}=\frac{{\bf p}^2}{2p^{+}}],\quad \\
&&[x^{-^{\prime }}=\frac{{\bf x}^2}2,\,\,p^{+^{\prime }}=0],\quad
[x^{+^{\prime }}=1,\,\,\,p^{-^{\prime }}=\left( {\bf x}\cdot {\bf p}%
-x^{-}p^{+}\right) ],\quad  \nonumber
\end{eqnarray}
The ones in the first line $[{\bf x,p]},[x^{-},p^{+}]$ are the true
canonical operators for the relativistic particle, which are quantized
according to the usual canonical rules 
\begin{equation}
\left[ {\bf x}^i{\bf ,p}^j\right] =i\delta ^{ij},\quad \left[
x^{-},p^{+}\right] =i\eta ^{+-}=-i.
\end{equation}
On the other hand, $x^{+}=0,\,\,x^{+^{\prime }}=1,\,p^{+^{\prime }}=0$ are
gauge choices and $p^{-},p^{-^{\prime }},x^{-^{\prime }}$ are dependent
operators which must be replaced by the given expressions in all gauge
invariant observables.

Recall that the Lorentz generators $L^{MN}$ are gauge independent and
commute with the Sp$\left( 2,R\right) $ generators. Therefore they can be
expressed in any gauge, consistently with the constraints, by simply
replacing our gauge choice (\ref{gauge1},\ref{gauge11}) into (\ref
{generators}). Thus, we obtain 
\begin{eqnarray}
L^{ij} &=&{\bf x}^{i}{\bf p}^{j}-{\bf x}^{j}{\bf p}^{i} \\
L^{+i} &=&-{\bf x}^{i}p^{+},\quad L^{-i}=x^{-}{\bf p}^{i}-\frac{{\bf p}^{j}%
{\bf x}^{i}{\bf p}^{j}}{2p^{+}} \\
L^{+-} &=&-\frac{1}{2}\left( x^{-}p^{+}+p^{+}x^{-}\right) ,\quad
L^{-^{\prime }+}=\frac{1}{2}{\bf x}^{2}p^{+} \\
L^{+^{\prime }+} &=&p^{+},\quad L^{+^{\prime }-}=\frac{{\bf p}^{2}}{2p^{+}}%
,\quad L^{+^{\prime }i}={\bf p}^{i} \\
L^{+^{\prime }-^{\prime }} &=&\frac{1}{2}\left( {\bf x}\cdot {\bf p+p}\cdot 
{\bf x}-x^{-}p^{+}-p^{+}x^{-}\right) \\
L^{-^{\prime }-} &=&\left[ 
\begin{array}{l}
\frac{1}{8p^{+}}\left( {\bf x}^{2}{\bf p}^{2}+{\bf p}^{2}{\bf x}^{2}-2\alpha
\right) \\ 
-\frac{x^{-}}{2}\left( {\bf x}\cdot {\bf p+p}\cdot {\bf x}\right)
+x^{-}p^{+}x^{-}
\end{array}
\right] \\
L^{-^{\prime }i} &=&\left[ 
\begin{array}{l}
\frac{1}{2}{\bf x}^{j}{\bf p}^{i}{\bf x}^{j}-\frac{1}{2}{\bf x}\cdot {\bf px}%
^{i} \\ 
-\frac{1}{2}{\bf x}^{i}{\bf p}\cdot {\bf x}+\frac{1}{2}{\bf x}^{i}\left(
x^{-}p^{+}+p^{+}x^{-}\right)
\end{array}
\right]
\end{eqnarray}
where operators are ordered to insure that all components of $L^{MN}$ are
hermitian. All possible ordering constants are uniquely fixed by hermiticity
except for the parameter $\alpha $ in $L^{-^{\prime }-}$. Our aim is to show
that these operators form the correct commutation rules for SO$\left(
d,2\right) $, namely 
\begin{equation}
\left[ L_{MN},L_{PQ}\right] =i\eta _{MP}L_{NQ}+i\eta _{NQ}L_{MP}-i\eta
_{NP}L_{MQ}-i\eta _{MQ}L_{NP}.  \label{sod2}
\end{equation}
This requirement fixes the parameter $\alpha =-1$. In particular 
\begin{equation}
\left[ L^{-^{\prime }i},L^{-\,j}\right] =i\delta ^{ij}L^{-^{\prime }-},\quad
\rightarrow \quad \alpha =-1.
\end{equation}
In a laborious calculation it can be verified that our construction
satisfies the correct commutation rules. The structure of the algebra may be
described as follows. First note that $L^{\mu \nu }=(L^{ij},L^{\pm
i},L^{+-}) $ form the SO$\left( d-1,1\right) $ Lorentz algebra, and that $%
p^{\mu }=\left( L^{+^{\prime }+},L^{+^{\prime }-},L^{+^{\prime }i}\right) $
are the generators of translations. The set $\left( L^{\mu \nu },p^{\mu
}\right) $ forms the Poincar\'{e} algebra ISO$\left( d-1,1\right) $ in the
massless sector $p^{2}=0$. The operators $K^{\mu }=\left( L^{-^{\prime
}+},L^{-^{\prime }-},L^{-^{\prime }i}\right) $ are the special conformal
transformations and finally $D=L^{+^{\prime }-^{\prime }}$ is the dilatation
operator.

It is also useful to note that the subset $\left( L^{\pm ^{\prime }\pm
},L^{\pm ^{\prime }\mp },L^{+^{\prime }-^{\prime }},L^{+-}\right) $ form the
algebra of SO$\left( 2,2\right) .$ Since SO$\left( 2,2\right) =SL\left(
2,R\right) _L\otimes SL\left( 2,R\right) _R$, it is convenient to identify
the SL$\left( 2,R\right) _L\otimes $SL$\left( 2,R\right) _R$ combinations as 
\begin{eqnarray}
G_2^L &=&\frac 12\left( L_{+^{\prime }-^{\prime }}+L_{+-}\right) ,\quad
G_0^L\pm G_1^L=L_{\pm ^{\prime }\pm }\,\,, \\
G_2^R &=&\frac 12\left( L_{+^{\prime }-^{\prime }}-L_{+-}\right) ,\quad
G_0^R\pm G_1^R=L_{\pm ^{\prime }\mp }.
\end{eqnarray}
which satisfy $\left[ G_a^L,G_b^R\right] =0$ and 
\begin{eqnarray}
\left[ G_0^{L,R},G_1^{L,R}\right] &=&iG_2^{L,R},\,\,\,\left[
G_0^{L,R},G_2^{L,R}\right] =-iG_1^{L,R}, \\
\left[ G_1^{L,R},G_2^{L,R}\right] &=&-iG_0^{L,R},
\end{eqnarray}
In our case we found the representation 
\begin{eqnarray}
G_2^L &=&\frac 14\left( {\bf x}\cdot {\bf p+p}\cdot {\bf x}\right) -\frac
12\left( x^{-}p^{+}+p^{+}x^{-}\right) \\
G_0^L+G_1^L &=&p^{+},\quad \\
G_0^L-G_1^L &=&\left[ 
\begin{array}{l}
\frac 1{8p^{+}}\left( {\bf x}^2{\bf p}^2+{\bf p}^2{\bf x}^2-2\alpha \right)
\\ 
-\frac{x^{-}}2\left( {\bf x}\cdot {\bf p+p}\cdot {\bf x}\right)
+x^{-}p^{+}x^{-}
\end{array}
\right]
\end{eqnarray}
and 
\begin{eqnarray}
G_2^R &=&\frac 14\left( {\bf x}\cdot {\bf p+p}\cdot {\bf x}\right) ,\quad \\
G_0^R+G_1^R &=&\frac{{\bf p}^2}{2p^{+}},\quad \\
G_0^R-G_1^R &=&\frac 12{\bf x}^2p^{+}.
\end{eqnarray}
These structures do indeed correctly form the Lie algebras of SO$\left(
2,2\right) =SL\left( 2,R\right) _L\otimes SL\left( 2,R\right) _R$. We
compute the quadratic Casimir operator of each $SL\left( 2,R\right) $ by $%
j\left( j+1\right) =G_0^2-G_1^2-G_2^2$. We find 
\begin{eqnarray}
j_R\left( j_R+1\right) &=&\frac 14{\bf L}^2+\frac 1{16}\left( d-2\right)
^2-\frac 14\left( d-2\right) \\
j_L\left( j_L+1\right) &=&j_R\left( j_R+1\right) -\frac{1+\alpha }4
\end{eqnarray}
where ${\bf L}^2=\frac 12L_{ij}L^{ij}$ is the Casimir operator for the
orbital rotation subgroup SO$\left( d-2\right) $ 
\begin{equation}
{\bf L}^2={\bf p}^i{\bf x}^2{\bf p}^i-{\bf p}\cdot {\bf x\,\,x}\cdot {\bf p.}
\end{equation}
Note that for $\alpha =-1$ we have $j_L=j_R$. The overall quadratic Casimir
for SO$\left( d,2\right) $ of eq.(\ref{c2}) takes the form 
\begin{eqnarray}
C_2 &=&\left\{ L^{+^{\prime }+},L^{-^{\prime }-}\right\} +\left\{
L^{+^{\prime }-},L^{-^{\prime }+}\right\} -\left( L^{+^{\prime }-^{\prime
}}\right) ^2-\left( L^{+-}\right) ^2  \nonumber \\
&&-\left\{ L^{+^{\prime }i},L^{-^{\prime }i}\right\} -\left\{
L^{+i},L^{-i}\right\} +\frac 12L_{ij}L^{ij} \\
&=&{\bf L}^2+\frac 14\left( d-2\right) ^2-\left( d-2\right) \\
&&-2{\bf L}^2-\frac 12\left( d-2\right) ^2+{\bf L}^2  \nonumber \\
&=&-\frac{d^2}4+1
\end{eqnarray}
As expected, the ``orbital'' part involving the canonical pairs $%
(x^{-},p^{+})$ and $({\bf x,p)}$ dropped out. By comparison to the covariant
form (\ref{c2}) we have verified that $C_2(Sp(2))=0$. This makes sense since
we have enforced the constraints at the classical level and thereby
guaranteed that the Sp$\left( 2,R\right) $ generators vanish in the physical
sector.

\subsection{Lorentz covariant quantization and field theory}

We may choose the gauge for the free particle partially to the following
form in the basis $M=\left( +^{\prime },-^{\prime },\mu \right) $ 
\begin{eqnarray}
X^M\left( \tau \right) &=&\left( 1,\frac{x^2\left( \tau \right) }2,x^\mu
\left( \tau \right) \right) ,  \label{covariantfix} \\
P^M\left( \tau \right) &=&\left( 0,\,p\left( \tau \right) \cdot x\left( \tau
\right) \,,p^\mu \left( \tau \right) \right) .  \nonumber
\end{eqnarray}
There remains the gauge degree of freedom that corresponds to $\tau $%
-reparametrization $\omega ^{22}\left( \tau \right) $ and the corresponding
constraint $p^2\left( \tau \right) =0$. The independent canonical pairs are
quantized as $\left[ x^\mu ,p^\nu \right] =i\eta ^{\mu \nu }$, which is
Lorentz covariant. Physical states $|\phi >$ must satisfy the $p^2|\phi >=0$
condition weakly. The well known solution may be given in $x$-space $\phi
\left( x\right) =<x|\phi >,$ where $<x|p_\mu =-i\frac \partial {\partial
x^\mu }<x|$, for which the constraint takes the form of the Klein-Gordon
equation for a massless particle 
\begin{equation}
\Box \phi \left( x\right) =0.
\end{equation}
The field theory ``effective action'' that gives this equation is 
\begin{equation}
S_{eff}=\frac 12\int d^dx\,\frac{\partial \phi }{\partial x^\mu }\frac{%
\partial \phi }{\partial x^\nu }\eta ^{\mu \nu }.  \label{seff}
\end{equation}
The solutions of the constraint are well known 
\begin{equation}
\phi \left( x\right) =\int \frac{d^dk\,\,\theta \left( k^0\right) }{\left(
2\pi \right) ^{d-1}}\,\,\delta \left( k^2\right) \,\left[ a\left( k\right)
\,e^{ik\cdot x}+a^{\dagger }\left( k\right) \,e^{-ik\cdot x}\right] .
\end{equation}
The states $|\phi >$ have the Lorentz invariant positive norm defined by 
\begin{eqnarray}
&<&\phi |\phi >=-\frac i2\int d^{d-1}x\,\left( \phi ^{*}\partial _0\phi
-\partial _0\phi ^{*}\phi \right) ,  \label{norm} \\
&=&\int d^{d-1}k\,\,\,\,a^{\dagger }\left( k\right) a\left( k\right) 
\nonumber
\end{eqnarray}
which is independent of the time component $x^0$ even though $x^0$ is not
integrated. The Lorentz invariance of this norm is well known from the study
of the Klein-Gordon equation, and can be seen by writing it in the form $%
\int dx\wedge \cdots \wedge dx\wedge \,J$ where $J_\mu =-\frac i2\left( \phi
^{*}\partial _\mu \phi -\partial _\mu \phi ^{*}\phi \right) $. If one wishes
one may rewrite the norm by choosing to fix the lightcone time $x^{+}$
instead of the ordinary time $x^0.$

Since SO$\left( d,2\right) $ is not manifest, we must check that the gauge
invariant conserved symmetry generators for SO$\left( d,2\right) $ have the
correct commutation rules (\ref{sod2}). We must first compute the gauge
invariant $L_{MN}$ in terms of $(x^\mu ,p^\mu )$ by inserting our gauge
choice and solutions of constraints given in (\ref{covariantfix}). We find

\begin{eqnarray}
L^{+^{\prime }-^{\prime }} &=&\frac 12\left( p\cdot x+x\cdot p\right) +i \\
L^{+^{\prime }\mu } &=&p^\mu \\
L^{-^{\prime }\mu } &=&\frac 12x_\lambda p^\mu x^\lambda -\frac 12x^\mu
p\cdot x-\frac 12x\cdot px^\mu -ix^\mu \\
L^{\mu \nu } &=&x^\mu p^\nu -x^\nu p^\mu
\end{eqnarray}
where operators are ordered. The commutation rules for SO$\left( d,2\right) $
are satisfied. All ordering ambiguities are uniquely determined by
hermiticity 
\begin{equation}
<\phi _1|L^{MN}\phi _2>=<L^{MN}\phi _1|\phi _2>,
\end{equation}
relative to the non-trivial Lorentz invariant norm in (\ref{norm}). This is
the reason for the appearance of the anomalous corrections proportional to $%
i $ in $L^{+^{\prime }-^{\prime }},L^{-^{\prime }\mu }$. Without these
anomaly pieces the generators are not hermitian. As a check that we have
correctly ordered our operators we compute the dimension of the scalar field
by applying $L^{+^{\prime }-^{\prime }}$ on it 
\begin{eqnarray}
iL^{+^{\prime }-^{\prime }}\phi \left( x\right) &=&<x|iL^{+^{\prime
}-^{\prime }}|\phi >  \nonumber \\
&=&\frac 12x\cdot \partial \phi +\frac 12\partial \cdot \left( x\phi \right)
-\phi \\
&=&x\cdot \partial \phi +\left( \frac d2-1\right) \phi \ .  \nonumber
\end{eqnarray}
The dimension ($\frac d2-1)$ is the correct dimension of the scalar field in
the effective field theory action (\ref{seff}). We also see that by
replacing $p_\mu =-i\partial /\partial x^\mu $ we arrive at the well known
construction of the conformal group in terms of differential operators as
known in field theory. The effective field theory $S_{eff}$ as well as the
dot product are invariants under these SO$\left( d,2\right) $ conformal
transformations applied on the field 
\begin{equation}
\delta \phi \left( x\right) =i\varepsilon _{MN}L^{MN}\phi \left( x\right)
,\quad \delta S_{eff}=0=\delta \left( <\phi _1|\phi _2>\right) .
\end{equation}
Thus $L^{+^{\prime }-^{\prime }}$ is the dimension operator, $L^{+^{\prime
}\mu }$ is the translation operator, $L^{-^{\prime }\mu }$ is the generator
of special conformal transformations and $L^{\mu \nu }$ is the generator of
Lorentz transformations.

We can now compute the quadratic Casimir operator for SO$\left( d,2\right) $
in this gauge. As we have argued in the previous section, its value is gauge
invariant, therefore it can be computed in any gauge. We find that it
reduces to a number 
\begin{eqnarray}
C_2 &=&-\left( L^{+^{\prime }-^{\prime }}\right) ^2-\left\{ L^{+^{\prime
}\mu },L^{-^{\prime }\nu }\right\} \eta _{\mu \nu }+\frac 12L_{\mu \nu
}L^{\mu \nu } \\
&=&-\frac{d^2}4\,+1.
\end{eqnarray}
where all $(x,p)$ dependence has dropped out\footnote{%
The dropping out of the orbital part is a phenomenon that occurs more
generally for any Casimir operator in a more general construction available
for {\it any group} \cite{ibsfets}. For example, a more general construction
for SO$\left( d,2\right) $ including the spin operator $s^{\mu \nu }$ and
the anomalous dimension operator $d_0$ 
\begin{eqnarray*}
J^{+^{\prime }-^{\prime }} &=&L^{+^{\prime }-^{\prime }}+id_0,\quad
J^{+^{\prime }\mu }=L^{+^{\prime }\mu },\quad \\
J^{-^{\prime }\mu } &=&L^{-^{\prime }\mu }-id_0x^\mu -s^{\mu \lambda
}x_\lambda ,\quad J^{\mu \nu }=L^{\mu \nu }+s^{\mu \nu }
\end{eqnarray*}
also has the property that {\it all} Casimir operators do not depend on the
``orbital'' operators $(x,p)$ contained in the $L^{MN}$. In particular the
quadratic casimir is $C_2=-\frac{d^2}4+\left( d_0+1\right) ^2+\frac 12s^{\mu
\nu }s_{\mu \nu }.$}. The value of the gauge invariant quadratic Casimir is
again the same$.$ This fixes the Sp$\left( 2,R\right) $ representation
uniquely to $C_2(Sp\left( 2\right) )=0$ in agreement with the previous
sections.

One may be puzzled by questions such as follows: Originally the operator $%
L^{+^{\prime }-^{\prime }}$ was a transformation that acted purely in the
extra dimensions $X^{\pm ^{\prime }}$ while leaving Minkowski space $x^\mu $
untouched; how can it now act like the scale transformations in Minkowski
space? The answer is that we chose the gauge $X^{+^{\prime }}=1$ that fixed
a scale. However linear transformation in global SO$\left( d,2\right) $
transforms $X^M$ out of this gauge slice. To come back to the same gauge one
must apply also a duality gauge transformation on $X^M\left( \tau \right) $.
The duality gauge transformation that corresponds to a rescaling of $%
X^{+^{\prime }}$ also rescales the rest of the components. This is precisely
what the operator $L^{+^{\prime }-^{\prime }}$ does on Minkowski space. The
structure of the {\it gauge invariant} operator $L^{+^{\prime }-^{\prime }}$
``knows'' that this gauge transformation must be performed on $X^M\left(
\tau \right) $.

Through our construction, the conformal group of massless field theories has
now acquired the new meaning of being the Lorentz-like group in an actual
space-time with two timelike dimensions $X^M$. The conformal field theory $%
S_{eff}$ has been expressed in a fixed gauge of the larger $d+2$ dimensional
space. There should exist a fully covariant effective field theory
corresponding to the SO$\left( d,2\right) \otimes Sp\left( 2,R\right) $
covariant quantization. The fully covariant action in $d+2$ dimensions would
collapse to the effective action of a massless particle given above upon
gauge fixing. Such a field theory may be formulated in terms of a bi-local
field $\phi (X_1^M,X_2^M)$.

\section{Outlook}

We have seen that the familiar free massless particle in $d$-dimensional
Minkowski space-time may be viewed as living in a larger space-time of $d+2$
dimensions. The higher space-time includes gauge degrees of freedom, but in
their presence the full SO$\left( d,2\right) $ conformal invariance takes
the new meaning of being the linear ``Lorentz symmetry'' in a space-time
that includes two timelike dimensions. Which of the two ``times'' $%
x^{0^{\prime }},x^0$ is {\it the} familiar time coordinate? For the gauge
choice we have made, time is $x^0$ and with it we have described the
dynamics a free particle. However, there are other choices of time as we
have demonstrated in the classical solutions here, and quantum solutions in
another paper \cite{ibdual}. For other choices of time the Hamiltonian is
different and the physics looks different (such as H-atom), even though we
are describing the same overall system that corresponds to a single unique
representation of the conformal group SO$\left( d,2\right) $. So, the
concept of ``time'' seems to be more general, and both of our two times play
a physical role. We may say that for the free massless particle the
appearance of {\it conformal symmetry {\bf is} the manifestation of a larger
space-time that includes two timelike coordinates}. Similarly, for the
H-atom and other dual systems, the presence of the conformal symmetry {\bf %
is }part of the evidence of the presence of two timelike dimensions.

Duality and the concept of two times are meshed together in our theory. The
duality we found is in the same spirit of the duality symmetry of M-theory,
but its realization requires two timelike dimensions in target space $%
X^M\left( \tau \right) $. This is more in line with the ideas of S-theory 
\cite{stheory} and F-theory \cite{ftheory}. In our case, we have actually
constructed an action for a miniature s-theory, which should serve as a
guide for constructing a full fledged S-theory in $\left( 10,2\right) $ and
perhaps even in $(11,3)$ dimensions \cite{14d}.

It may be interesting to view our theory as {\it conformal gravity} on the
worldline as noted earlier in the paper. We may then regard the gauge fields 
$\left( A^{22},A^{12},A^{11}\right) $ as the gauge fields for translations,
dilatations, and special conformal transformations respectively. Our theory
may be used as a guide for generalizations from the worldline to the
worldsheet or worldvolume for various $p$-branes. Although conformal gravity
on the worldsheet has been considered before \cite{schoutens}, our approach
in phase space is somewhat different and may yield a new and different
action. Such a reformulation of $p$-brane actions would permit the
introduction of two timelike dimensions in $X^{M}(\tau ,\sigma _{1},\cdots
,\sigma _{p})$ just as in the particle case $p=0$.

The present paper, as well as some of our previous papers \cite{ibkounnas}-%
\cite{sparticles}, are attempts to take the concept of two or more timelike
dimensions seriously. We may ask: are there more observable effects of two
timelike dimensions besides the conformal invariance and the duality
connections we have suggested? To answer such questions it would be useful
to study interacting theories that are consistent with the gauge duality
symmetries. This is essential in order to avoid ghosts. As a first step one
may explore the interacting theory $S_G$ that would result from a curved
background in $\left( d,2\right) $ dimensions. This is formulated by taking
a curved metric $G_{MN}\left( X_1,X_2\right) $ instead of $\eta _{MN}$ in
the action (\ref{action}). One way to maintain the local Sp$\left(
2,R\right) $ symmetry is to take $G_{MN}$ as a function of only the gauge
invariant combination $X_i^MX_j^N\varepsilon ^{ij}$. It is also possible to
study interactions using $S_A$ in the presence of background gauge fields $%
A_M^i\left( X_1,X_2\right) $ that couple in a gauge invariant way to $D_\tau
X_i^M\left( \tau \right) $ in $\left( d+2\right) $ dimensions. Here it would
be interesting to explore the possible relation between our Sp$\left(
2\right) $ doublet $A_M^i$ and the electric-magnetic dual potentials of
Maxwell's theory and its generalizations \cite{seibwitt}. One thing that is
becoming clearer is that bi-local fields $\phi \left( X_1,X_2\right) $ are
probably going to be very useful for writing down the low energy effective
theories consistently with the local Sp$\left( 2,R\right) $ invariance.

The idea of bi-local fields also emerged before as a means of displaying the
hidden timelike dimensions in certain BPS sectors which provide short
representations of the superalgebra of S-theory \cite{stheory}. It was
emphasized that such BPS sectors, which reveal extra timelike dimensions in
black holes \cite{sentropy}, must be considered dual sectors to other BPS
solutions of M-theory. Progress along these and other directions for
interacting theories will be reported in the future. We hope that such
interacting theories would provide the means to discuss how to probe the
higher hidden dimensions and perhaps find some additional measurable
consequences and tests.

We would like to think that the presence of duality \cite{mtheory} and
conformal symmetry \cite{maldacena} in M-theory, as well as in special super
Yang-Mills theories under current consideration, are also signs of the
presence of higher dimensions, and in particular of extra timelike
dimensions. Indeed various signs that the mysterious theory may actually
have 12 dimensions with signature $\left( 10,2\right) $ has been
accumulating. It has also been argued that a fundamental theory that is {\it %
manifestly covariant} under both duality and supersymmetry requires 14
dimensions with signature $\left( 11,3\right) $ to display the covariance
(in the spirit of the current paper), and it must have certain ``BPS''
constraints that are due to gauge invariances \cite{14d}. The various ideas
outlined in this paper may be regarded as a small step toward a formulation
of such a theory.


\medskip

\begin{thebibliography}{99}
\bibitem{ibdual}  I. Bars, ``Conformal symmetry and duality between free
particle, harmonic oscillator and H-atom'', in preparation.

\bibitem{duff}  M. Duff and M.P. Blencowe, Nucl. Phys. {\bf B310} (1988) 387.

\bibitem{ibtokyo}  I. Bars, ``Duality and hidden dimensions'',
hep-th/9604200, lecture in the proceedings of the conference {\it Frontiers
in Quantum Field Theory}, Toyonaka, Japan, Dec. 1995; I. Bars Phys. Rev. 
{\bf D54 }(1996) 5203.

\bibitem{mtheory}  E. Witten, Nucl. Phys. {\bf B443} (1995) 85.

\bibitem{ftheory}  C. Vafa, Nucl. Phys. {\bf B469} (1996) 403.

\bibitem{stheory}  I. Bars, Phys. Rev. {\bf D55} (1997) 2373; I. Bars,
``Algebraic Structures in S-Theory'', hep-th/9608061, lectures in Second
Sakharov conf. 1996, and Strings-96 conf.

\bibitem{14d}  I. Bars, Phys. Lett. {\bf B403} (1997) 257.

\bibitem{martinec}  D. Kutasov and E. Martinec, Nucl. Phys. {\bf B477}
(1996) 652; Nucl. Phys. {\bf B477} (1196) 675; E. Martinec, ``Geometrical
structures of M-theory'', hep-th/9608017.

\bibitem{sezgin}  H. Nishino and E. Sezgin, Phys. Lett. {\bf B388} (1996)
569; E. Sezgin, Phys. Lett. {\bf B403 (}1997) 265; H. Nishino,
hep-th/9710141.

\bibitem{ibkounnas}  I. Bars and C. Kounnas, Phys. Lett. {\bf B402} (1997)
25; Phys. Rev. {\bf D56} (1997) 3664.

\bibitem{spartstring}  I. Bars and C. Deliduman, Phys. Rev. {\bf D56} (1997)
6579.

\bibitem{sparticles}  I. Bars and C. Deliduman, ``Gauge principles for
multi-superparticles'', hep-th/9710066.

\bibitem{sezrudy}  I. Rudychev and E. Sezgin, ``Superparticles in $D>11$
Revisited'', hep-th/9711128.

\bibitem{seibwitt}  N. Seiberg and E. Witten, Nucl. Phys. {\bf B426} (1994)
19.

\bibitem{hulltown}  C.M. Hull and P.K. Townsend, Nucl. Phys. {\bf B438 (}1995%
{\bf ) }109.

\bibitem{osp12}  I. Bars, ``Gauged duality and supersymmetry'', in
preparation.

\bibitem{marnelius}  R. Marnelius, Phys. Rev. {\bf D20 }(1979){\bf \ }2091.

\bibitem{siegel}  W. Siegel, Int. J. Mod. Phys. {\bf A3 }(1988){\bf \ }2713.

\bibitem{martensson}  U. Martensson, Int. J. Mod. Phys. {\bf A8 }(1993){\bf %
\ }5305.

\bibitem{followup}  S.M. Kuzenko and J.V. Yarevskaya, Mod. Phys. Lett. {\bf %
A11 }(1996){\bf \ }1653.

\bibitem{barut}  A. O. Barut and L. Bornzin, J. Math. Phys. {\bf 12} (1971)
841.

\bibitem{ibsfets}  I. Bars and K. Sfetsos, Nucl. Phys. {\bf B371} (1992) 507.

\bibitem{schoutens}  J. W. van Holten, Nucl. Phys. {\bf B277} (1986) 429, T.
Uematsu, Phys. Lett. {\bf B183} (1987) 304; K. Schoutens Nucl. Phys. {\bf %
B292} (1987) 151.

\bibitem{sentropy}  I . Bars, Phys. Rev. {\bf D55 }(1997) 3633.

\bibitem{maldacena}  J. Maldacena, ``The large N limit of superconformal
field theories and supergravity'', hep-th/9711200.
\end{thebibliography}
\end{document}